\documentclass[a4paper,11pt]{article}
\usepackage{pos}

\title{Thirty years of the Beauty Conference}

\author*[a]{Neville Harnew}

\affiliation[a]{University of Oxford,\\
  Denys Wilkinson Building, 1 Keble Road, Oxford OX1 3RH, UK}

\emailAdd{neville.harnew@physics.ox.ac.uk}

\abstract{This paper remembers thirty years of the Beauty conference series and celebrates its 20$^{\rm th}$  meeting. The conference highlights  are reviewed.
}

\FullConference{20th International Conference on B-Physics at Frontier Machines (Beauty2023)\\
 3-7 July, 2023\\
Clermont-Ferrand, France\\}

\begin{document}
\maketitle

\section{Introduction}

This paper highlights 30 years of the Beauty Conference series from a personal perspective. It is broken down into four distinct periods. In the early years, 1993--1999,  the predominant focus of the conference was to prepare the first dedicated $B$-physics experiments for future hadron machines.
The second period, 1999--2009, saw the first measurements of CP-violation and rare $B$ decays, dominated by the $B$-factory $e^+e^-$ and Tevatron experiments.
Between 2009--2022 came the LHC era, where the  ATLAS, CMS and LHCb experiments took the lead in $B$-physics measurements. Finally, beyond 2019 saw the start-up of the Belle-II experiment and the transition to the LHC upgrades, taking $B$-physics into a new high-luminosity era.  An overview of the timeline is presented in Fig.\,\ref{Fig:timeline}. A full list of the twenty Beauty conferences is given below.

\begin{figure} 
  \begin{center}
    \includegraphics[width=0.55\linewidth]{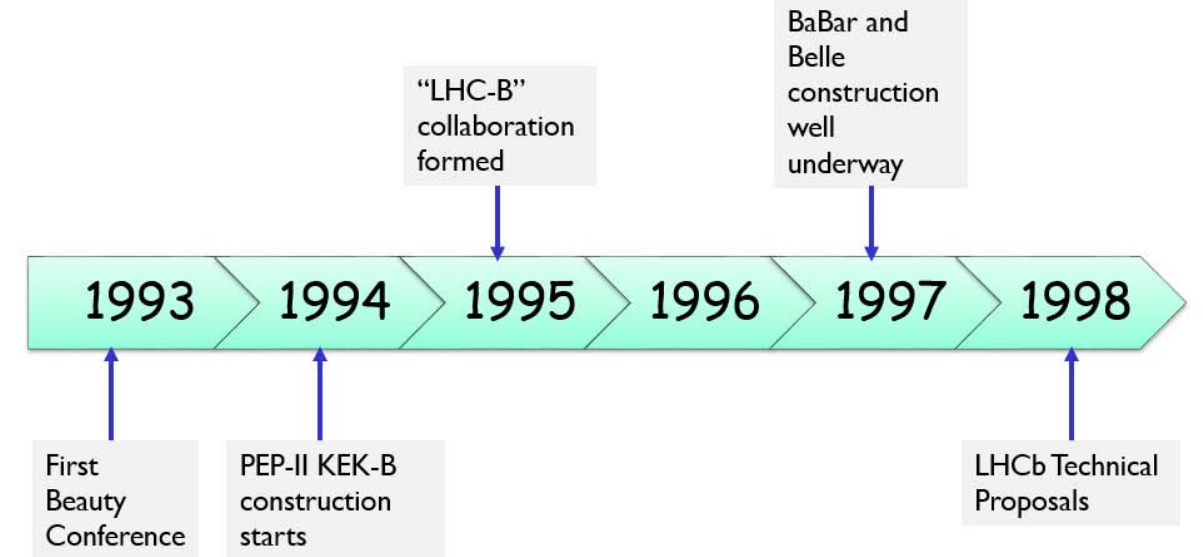} 
        \includegraphics[width=0.75\linewidth]{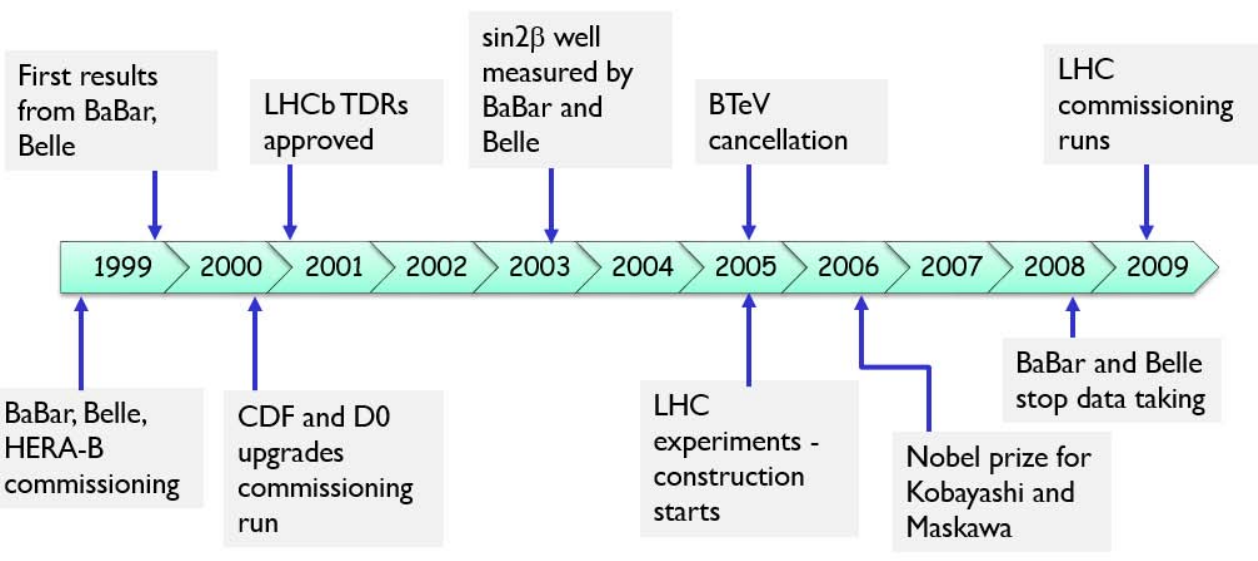}
            \includegraphics[width=0.85\linewidth]{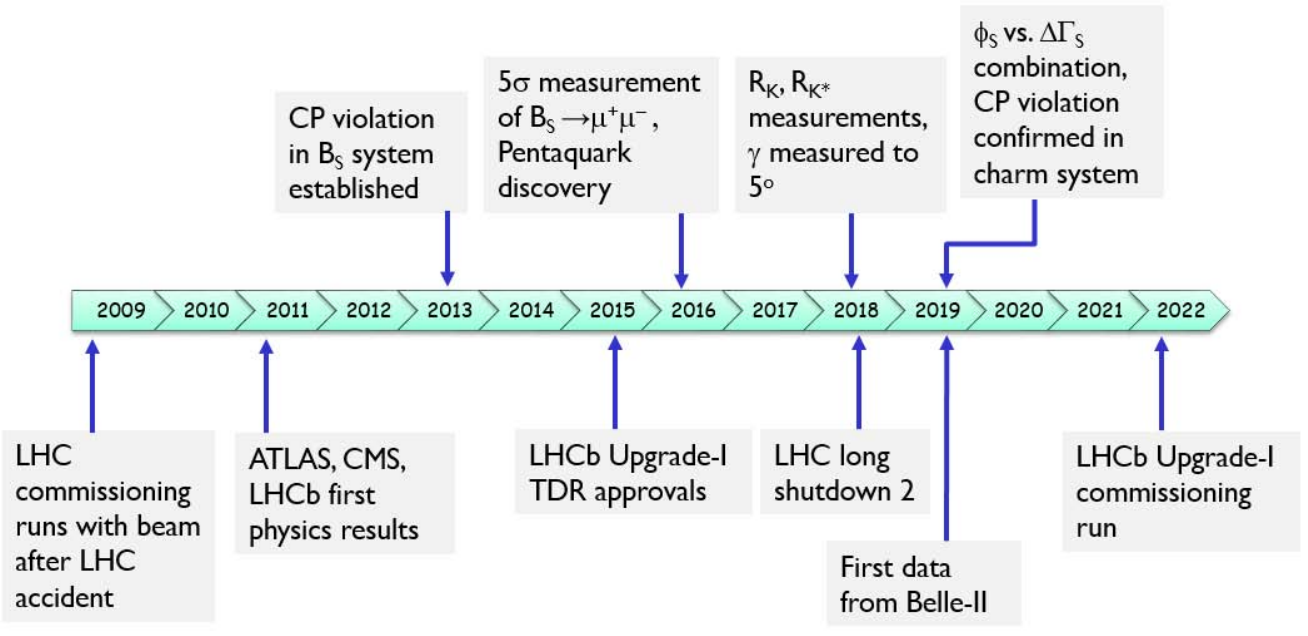} 
                \includegraphics[width=0.7\linewidth]{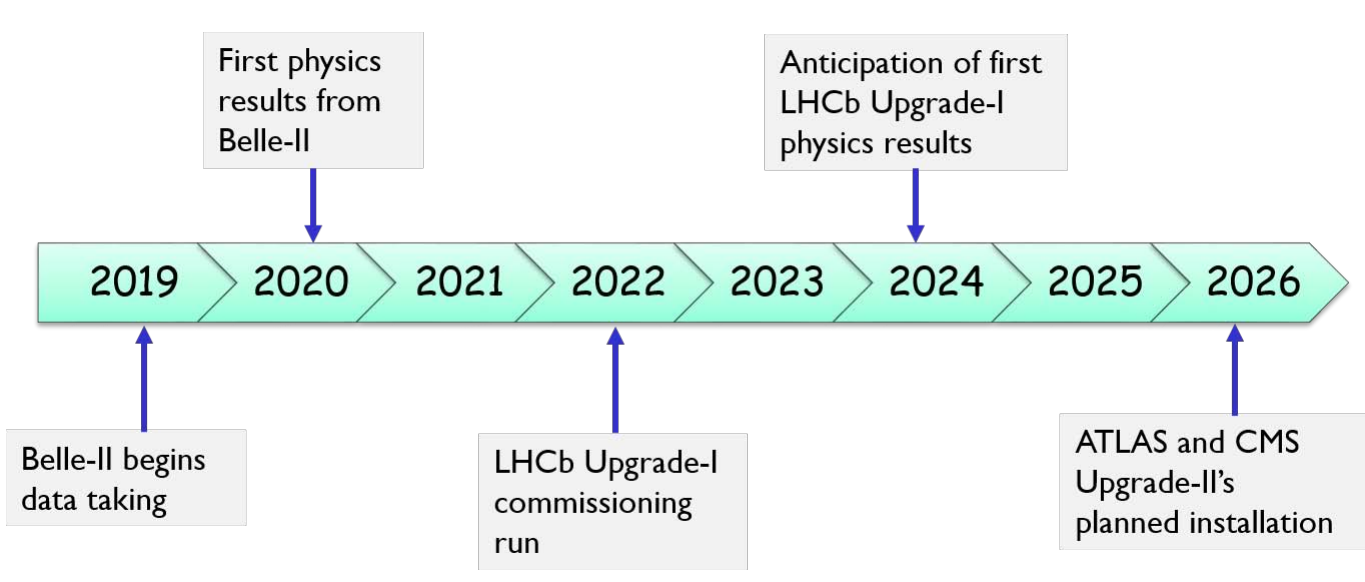} 
  \end{center}
  \caption{
From the top : the early years of the conference, 1993--1999;
the $e^+e^-$ and Tevatron era,  1999--2009; the LHC era, 2009--2022;  the era of Belle-II and the LHC upgrades, beyond 2019.
    }
  \label{Fig:timeline}
\end{figure}


\begin{center}
\begin{tabular}{  c c }
 {\bf Year} & \bf{Conference location} \\ 
 1993 & Liblice Castle, Melnik, Czech Republic \\
 1994 & Le Mont Saint Michel, Normandy, France \\
 1995 & Oxford, United Kingdom \\
 1996  & Rome, Italy \\
 1997 & Santa Monica, CA, United States \\
 1999 & Bled, Slovenia \\
 2000 & Sea of Galilee, Kibbutz Maagan, Israel \\
 2002 & Santiago de Compostela, Spain \\
 2003 & Pittsburgh, PA, United States \\
 2005 & Assisi, Perugia, Italy \\
 2006 & Oxford, United Kingdom \\
 2009 & Heidelberg, Germany \\
 2011 & Amsterdam, Netherlands \\
 2013 & Bologna, Italy \\
 2014  & Edinburgh, United Kingdom \\
 2016 & Marseille, France \\
 2018 & La Biodola, Elba Island, Italy \\
 2019 & Ljubljana, Slovenia \\
 2020 &  Kavli Institute, IPMU, Japan (online conference) \\
 2023 & Clermont-Ferrand, France \\ 
\end{tabular}
\end{center}

\section{The early years of the conference : 1993--1999}

\begin{figure}[h]
  \begin{center}
    \includegraphics[width=0.85\linewidth]{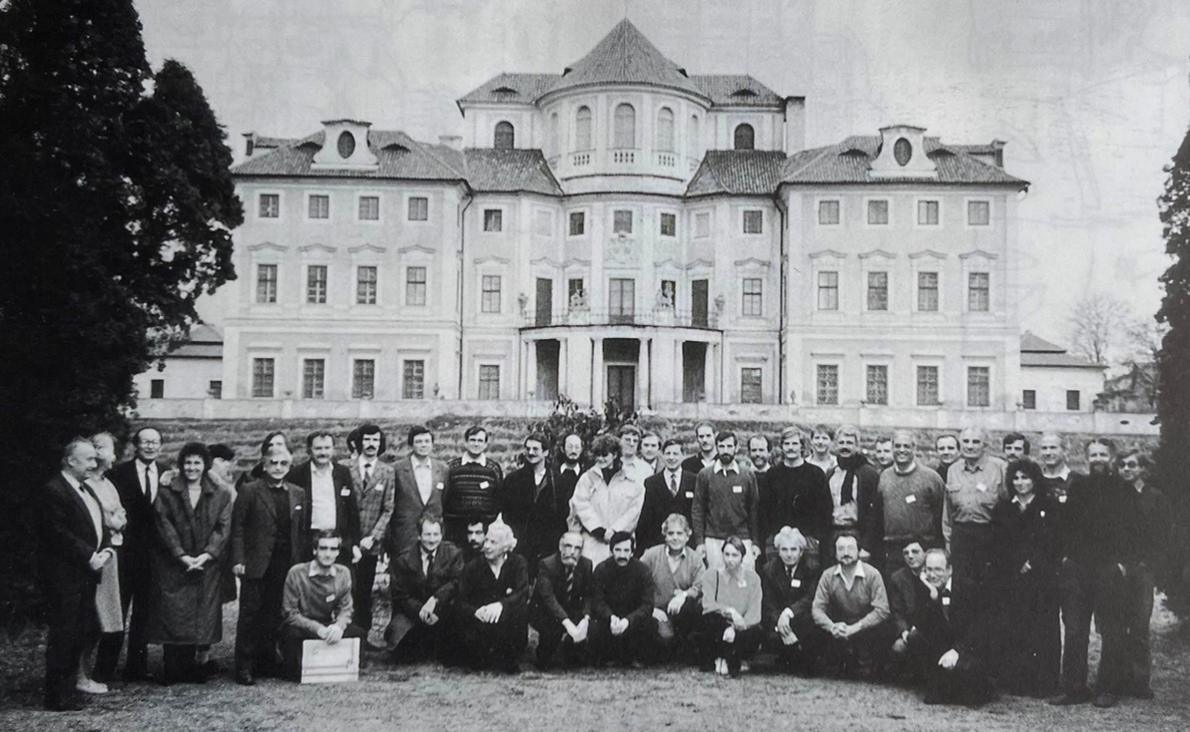} 
  \end{center}
  \caption{
A group photo of the participants of first Beauty Conference held in Liblice Castle, Melnik, Czech Republic. Peter Schlein can be seen kneeling front row, sixth from the right. 
    }
  \label{Fig:conference-photo}
\end{figure}

The first Beauty conference, Beauty 1993, was held at Liblice Castle, Melnik, Czech Republic~\cite{Beauty1993}, see Fig.\,\ref{Fig:conference-photo}. It was initiated by the ``father'' of the Beauty conference series, Peter Schlein (1932-2008) and close colleagues.
The conference began as a forum for discussion  of the merits of the different methods of $B$-physics experimentation, at that time 
at the $e^+e^-$ $B$-factories, the  LHC and the SSC accelerators, and with the HERA-B experiment.

\begin{figure}[h]
  \begin{center}
    \includegraphics[width=0.75\linewidth]{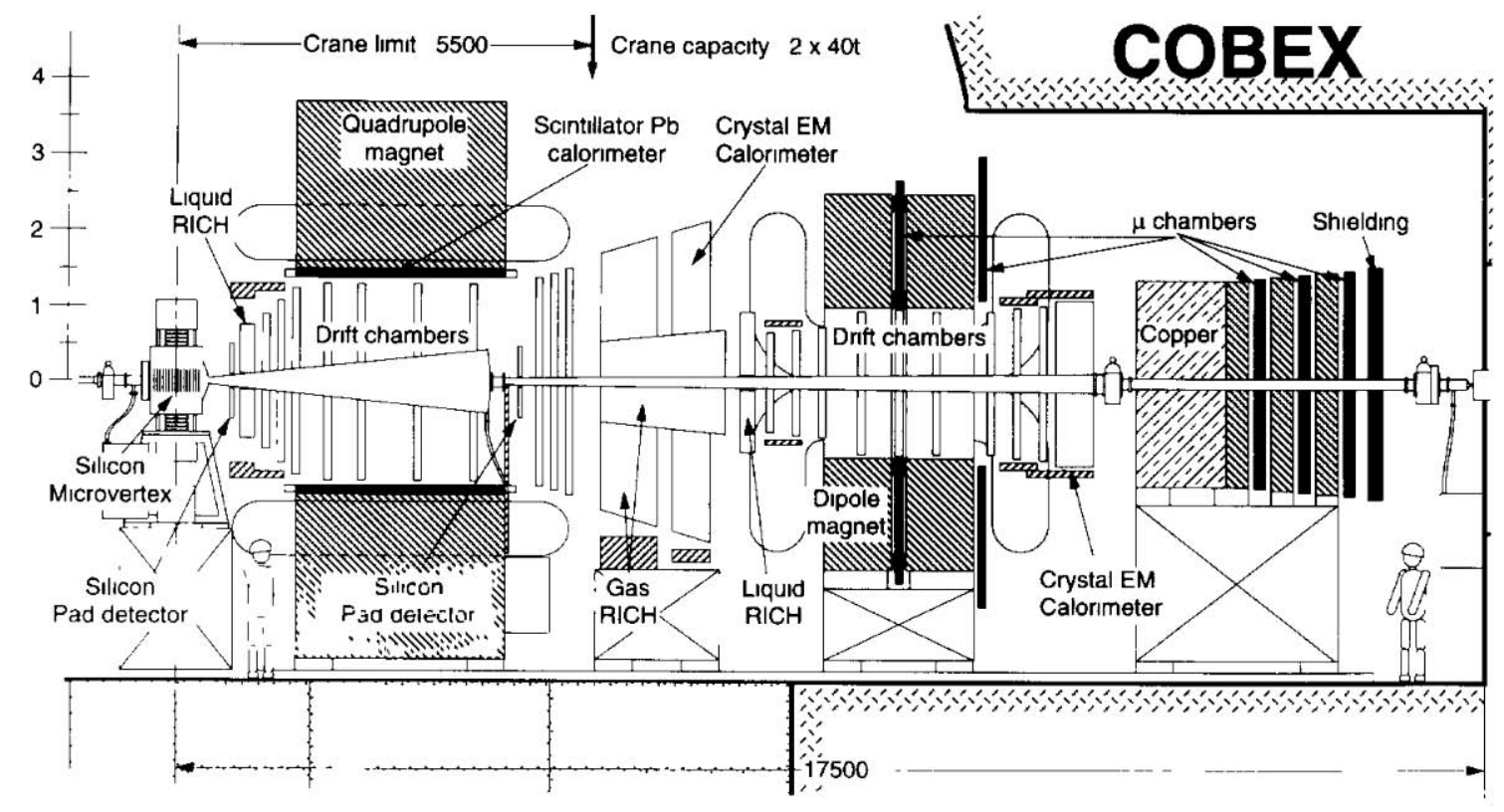} 
 
    \vspace{0.5truecm}
        \includegraphics[width=0.99\linewidth]{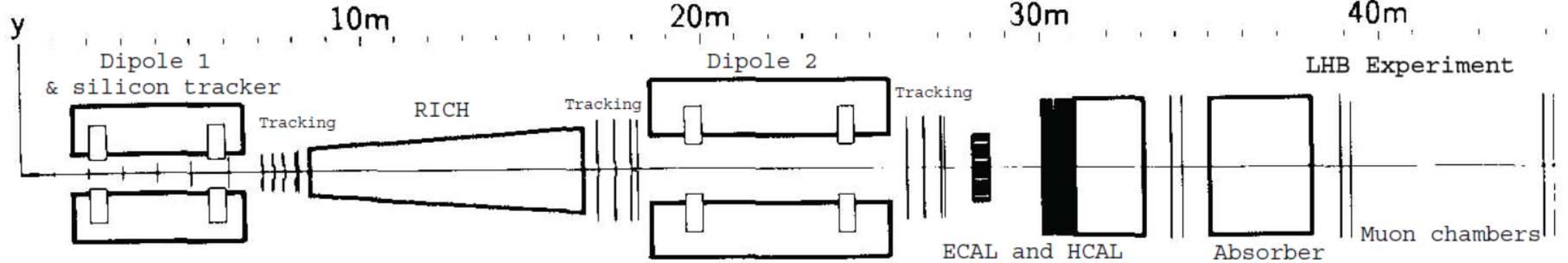}
        
            \vspace{0.5truecm}
            \includegraphics[width=0.75\linewidth]{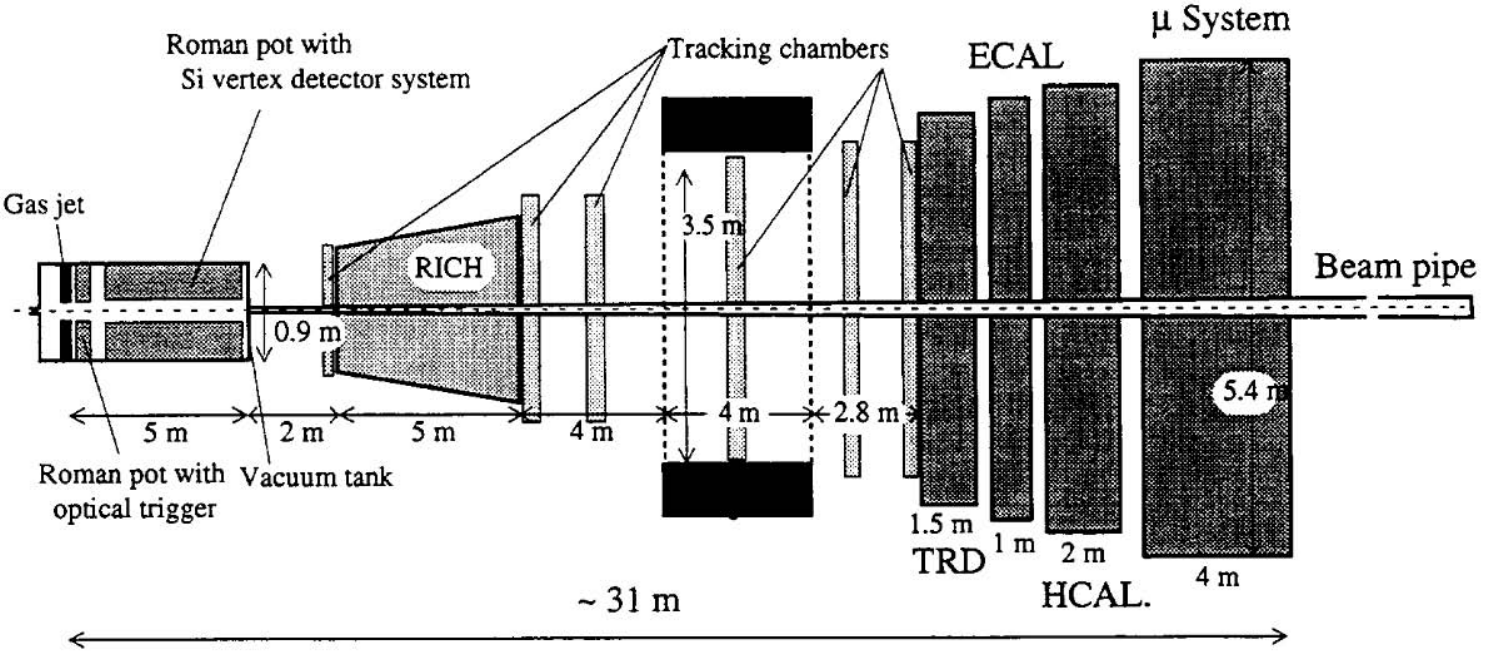} 
  \end{center}
  \caption{
The three propoals for  $B$ physics at the LHC: 
from the top, COBEX (LHC Collider mode),
LHB  (beam extraction),
GAJET (internal gas jet target). 
    }
  \label{fig:LHCC}
\end{figure}

The second conference, Beauty 1994, was held in  Mont-Saint-Michel, France\,\cite{Beauty1994}, and was arguably one of the most vibrant of the whole series.  Three proposals for $B$ physics at the LHC were presented and are shown in Fig.\,\ref{fig:LHCC}: 
COBEX, running in LHC collider mode,
the Large Hadron Beauty (LHB) experiment, running with beam extraction, 
and GAJET, running with an internal gas jet target. Emotions ran high in discussing the proposals' relative benefits, and 
in addition there was a  strong lobby for HERA-B.
After the conference, the CERN Large Hadron Collider Committee (LHCC) reviewed the strengths and weaknesses of the three proposals. The conclusions drawn were that  
none of the collaborations had the necessary resources,  
the collider mode had the greater potential, however 
an optimized design of spectrometer did not yet exist.
The Committee therefore encouraged all participants to join together to prepare a  Letter of Intent for a new collider-mode $B$-physics experiment.  The new collaboration was subsequently born and the ``LHC-B'' Letter of Intent was submitted\,\cite{LHCB-LoI}.  The collaboration was cemented at Beauty 1995 in Oxford\,\cite{Beauty1995}.

By the time of the Beauty 1996 conference in Rome\,\cite{Beauty1996}, the BaBar and Belle experiments were well in preparation.
The LHC-B experiment was pushing towards a Technical Proposal and  BTeV,  at the Tevatron, had received its initial approval. 
CDF reported the first $B^0\to J/\psi K^0_s$ signal to be observed at a hadron collider.

\section{The $e^+e^-$ $B$-Factory and Tevatron era : 1999--2009}


At Beauty 1999 in Bled, Slovenia\,\cite{Beauty1999}, the Belle, BaBar and  HERA-B experiments were commissioning and producing first results. The 
OPAL and CDF experiments provided the first hints of CP violation in the $B$ sector by measuring non-zero $\sin 2\beta~ (\sin 2\phi_1 )$, albeit with very limited precision: $\sin 2\beta = 3.2 \pm 1.9 \pm 0.5$ (OPAL) and $\sin 2\beta = 0.79 \pm 0.44$ (CDF). 
The Beauty conference then had a four-year gap. 

\begin{figure}[h]
  \begin{center}
        \includegraphics[width=0.49\linewidth]{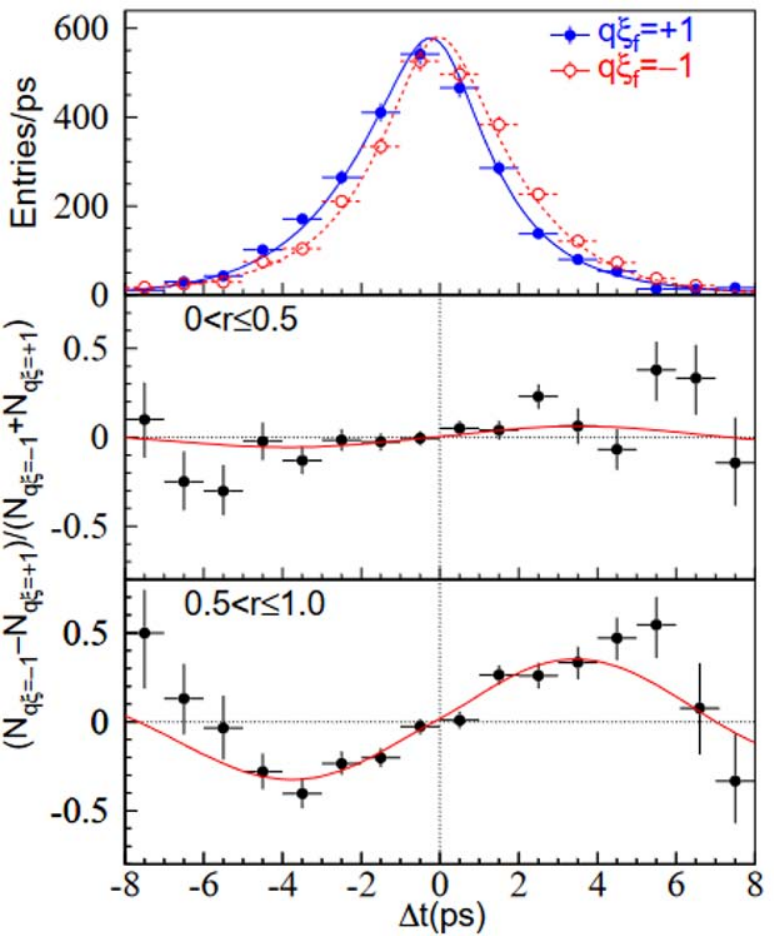}    \includegraphics[width=0.49\linewidth]{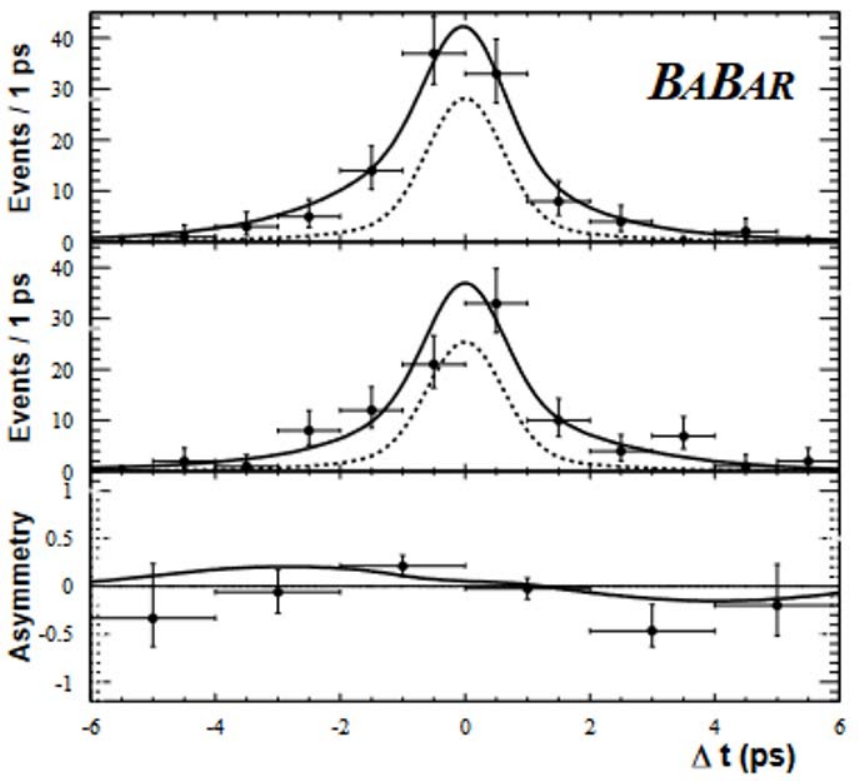}  
  \end{center}
  \caption{
Decay-time distributions and charge asymmetries and a function of decay time  for  (left) $B^0\to J/\psi K^0_s$ from Belle\,\cite{Bellephi1}, and (right)  $B^0\to \pi^+\pi^-$ from BaBar\,\cite{Babarphi2}. 
    }
  \label{Fig:sin2beta}
\end{figure}

By the time of Beauty 2003\,\cite{Beauty2003}, hosted at Pittsburgh, $\sin 2\beta~ (\sin 2\phi_1 ) $ had been well established by BaBar and Belle with better than a 10\% uncertainty from each experiment. Figure\,\ref{Fig:sin2beta}\,(left) shows the  Belle time-dependent asymmetry measurement in $B^0\to J/\psi K^0_s$ decays after 140\,fb$^{-1}$ of integrated luminosity, giving $\sin 2\phi_1 = 0.733 \pm 0.057 \pm 0.028$\,\cite{Bellephi1}. The first measurements of the CKM angles $\alpha~ (\phi_2 )$ and $\gamma~ (\phi_3 )$ were also emerging\,\cite{Babarphi2}, see Fig.\,\ref{Fig:sin2beta}\,(right).
 CDF and D0 were setting the standards for $B$ physics at hadron machines, measuring 5\% precision on $B$ lifetimes.
Unfortunately  HERA-B’s physics output was very limited, restricted to the measurement of the $b\overline{b}$ cross-section at 920\,GeV proton energy\,\cite{HERA-B}. 

At Beauty 2005 in Assisi, Italy\,\cite{Beauty2005}, BaBar  presented a new  measurement of angle $\alpha (\phi_2 ) = (103\substack{+9 \\ -11})^\circ$. 
The LHC experiment R\&D was now at an end, and construction had  started. 
Sadly this year had seen the cancellation of BTeV  by the US Department of Energy. 

\begin{figure}[h]
  \begin{center}
    \includegraphics[width=0.49\linewidth]{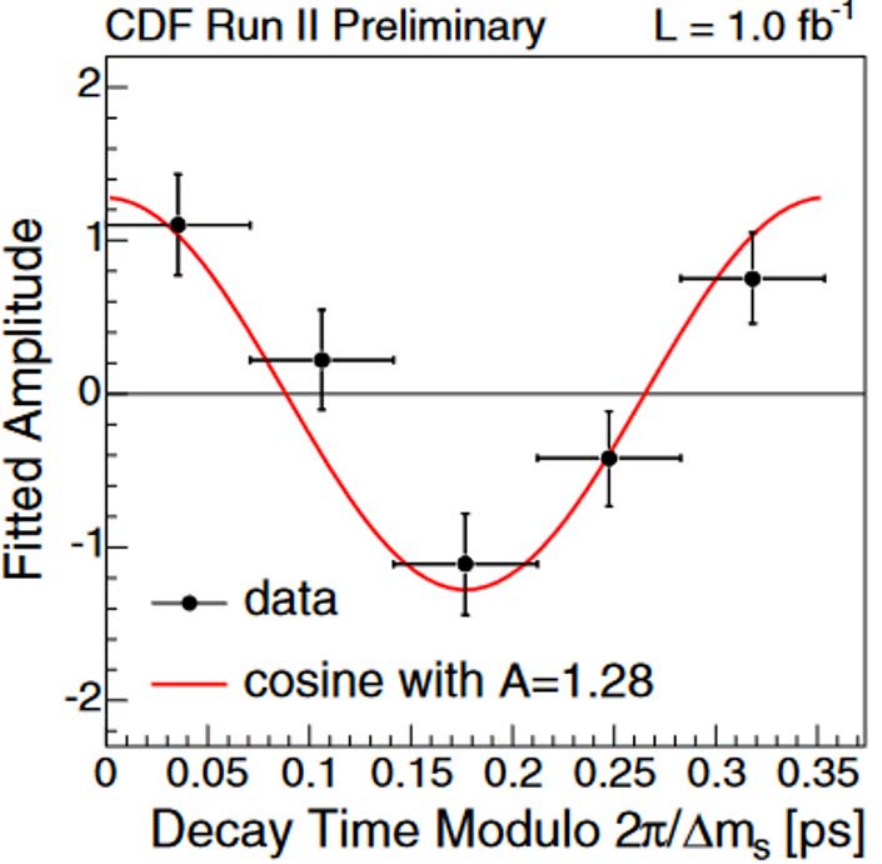} 
        \includegraphics[width=0.49\linewidth]{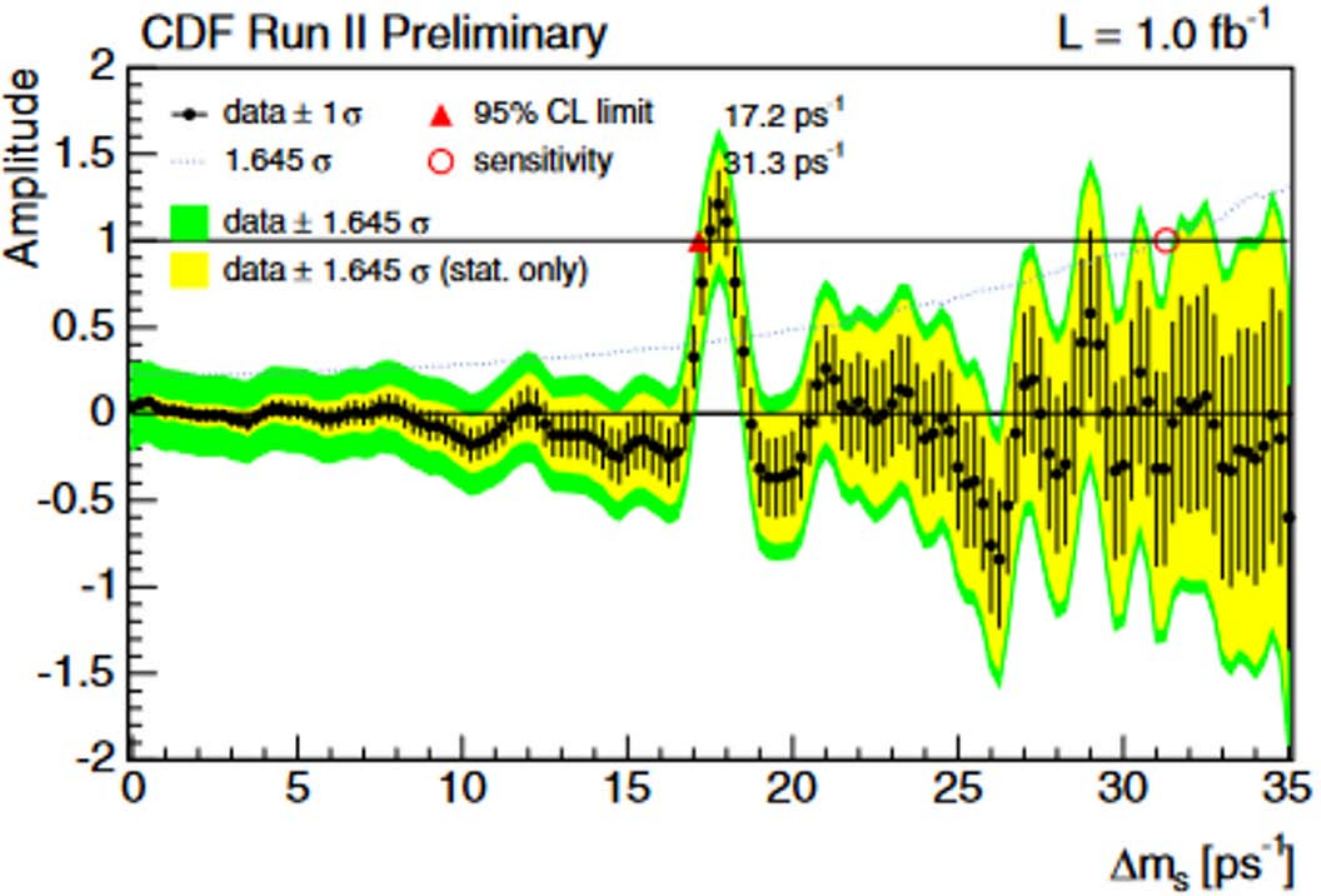} 
  \end{center}
  \caption{
The observation of $B_s - \overline{B}_s$  oscillations at the CDF experiment~\cite{CDFoscillations}. (Left) the oscillation signal in bins of proper decay time modulo the measured oscillation period,  (right) the $\Delta m_s$ amplitude scan. 
    }
  \label{Fig:CDFoscillations}
\end{figure}

The following year the conference re-visited Oxford with Beauty 2006\,\cite{Beauty2006}. BaBar and Belle made the first measurements of angle $\gamma (\phi_3 )$ in $B^+\rightarrow D^{(*)}K^{+(*)}$ decays, albeit with  30\% errors. Arguably the highlight of the conference was the measurement of the $B_s - \overline{B}_s$ oscillation frequency with significance greater than $5\sigma$ by CDF\,\cite{CDFoscillations}. The $\Delta m_s$ amplitude scan, shown in Fig.\,\ref{Fig:CDFoscillations}  yielded a value of $\Delta m_s = (17.77 \pm 0.10  \pm 0.07)$\,ps$^{-1}$. 
A quote from the conference editorial stated ``As this (in 2006) is the last conference in the series before the start-up of the LHC, Beauty 2006 has been a timely opportunity to review the status of the field.'' This statement was made prior to the LHC accident in 2008, which resulted in an additional year's gap in the conference programme.

\section{The LHC era : 2009--2022}

\begin{figure}[h]
  \begin{center}
    \includegraphics[width=0.5\linewidth]{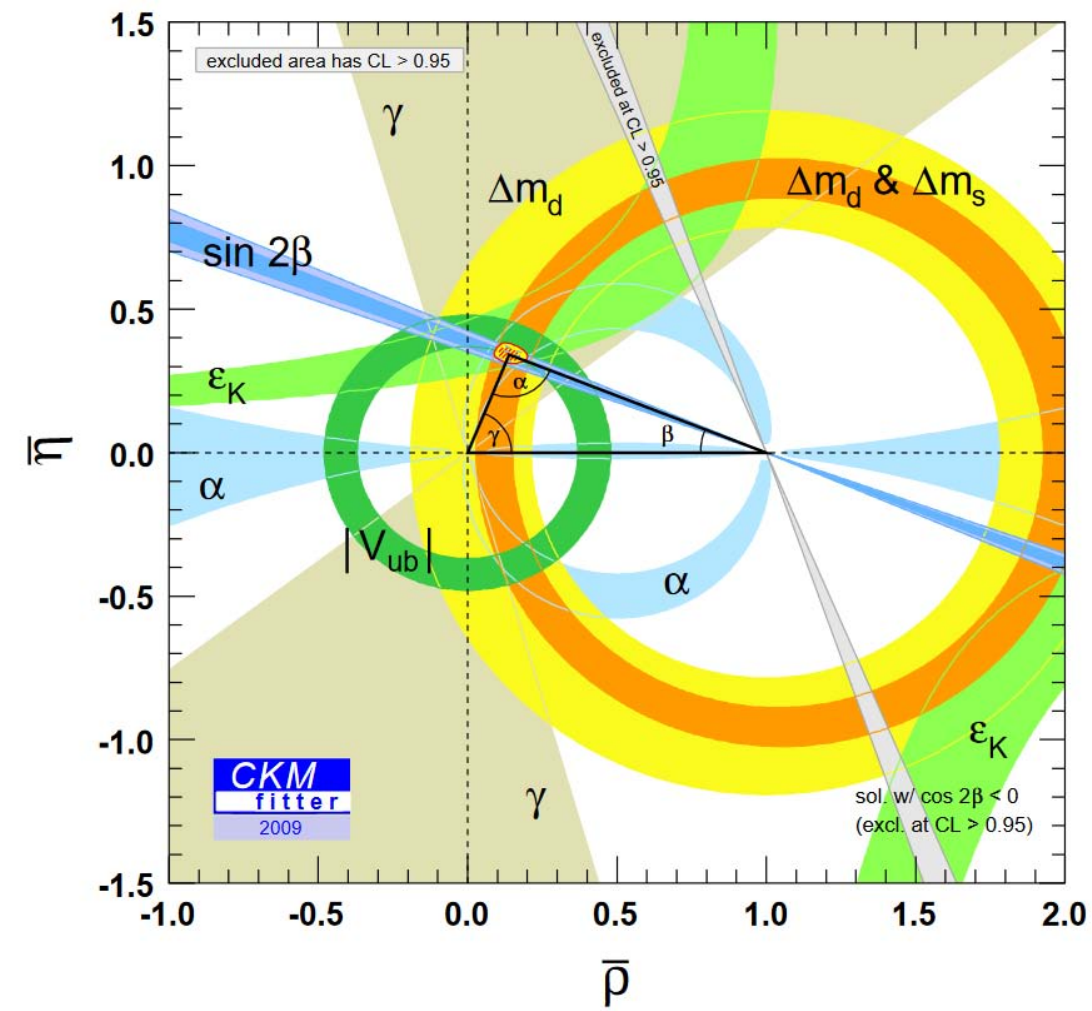} 
  \end{center}
  \caption{
The Unitarity Triangle in 2009\,\cite{CKMfitter}. Many of the measured parameters are dominated by BaBar and Belle measurements.
    }
  \label{fig:UT2009}
\end{figure}

Beauty 2009 in Heidelberg\,\cite{Beauty2009}  saw the first preliminary data from the LHC experiments after the LHC accident. Whilst statistics were limited, first $B$ lifetime measurements were presented. The conference also saw the end of data taking for BaBar and Belle  with a $\sim$1.5\,ab$^{-1}$  combined total integrated luminosity. An amazing legacy of $B$-factory results was presented, including 
the observation of $B\rightarrow\tau\nu$, the forward-backward asymmetry in $B\rightarrow K^*\ell^+\ell^-$, angle $\beta (\phi_1)$ now known to $1^\circ$, $\alpha (\phi_2)$ known to $5^\circ$ and $\gamma (\phi_3)$ known to better than $15^\circ$. In 2009 the unitarity triangle had been impressively constrained  by the $B$-factory experiments,  demonstrated in Fig.\,\ref{fig:UT2009}.

The first significant physics results from the LHC experiments were reported at Beauty 2011 in Amsterdam\,\cite{Beauty2011} . This included LHCb's observation of direct CP violation in $B\rightarrow K^+\pi^-$ decays  and the first observation of  $B_s\rightarrow J/\psi\phi$ decays from three LHC experiments, following first results from the Tevatron in 2009-2011. 

\begin{figure}[h]
  \begin{center}
    \includegraphics[width=0.49\linewidth]{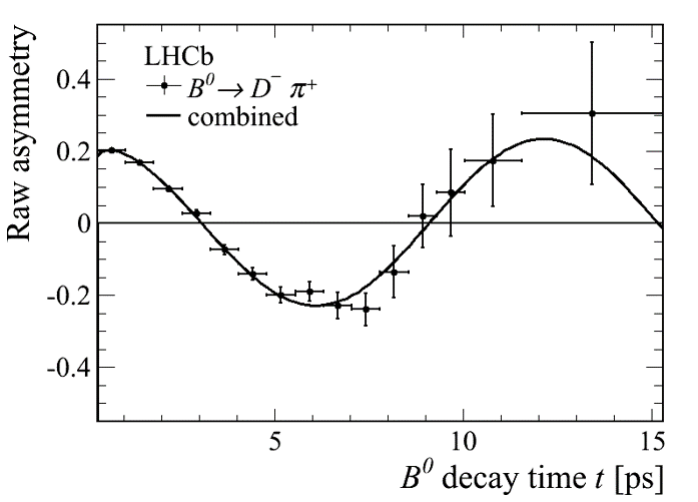} 
        \includegraphics[width=0.49\linewidth]{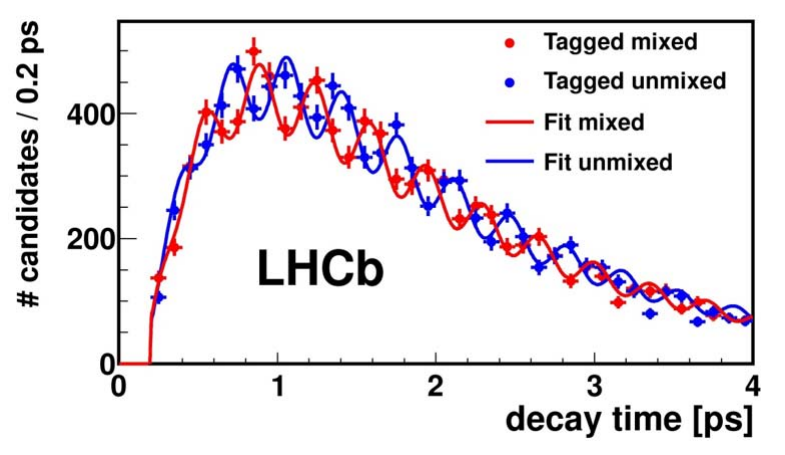} 
  \end{center}
  \caption{
 LHCb $B$ mixing measurements showing the $B-\overline{B}$ asymmetry versus $B$ decay time; (left) $B^0$
\,\cite{LHCbbmixing} and (right) $B^0_s$\,\cite{LHCbbsmixing}.}
  \label{Fig:mixing}
\end{figure}

Beauty 2013 in Bologna\,\cite{Beauty2013} saw the LHC experiments start to push the boundaries. This included the first evidence for $B_s\rightarrow \mu^+\mu^-$ from LHCb with a measured branching ratio of $(3.2\pm\substack{+1.4 \\ -1.2}\pm\substack{+0.5 \\ -0.3})\times 10^{-9}$. There were much-improved constraints on the $B_s$ mixing phase in $B_s\rightarrow J/\psi\phi$ decays from ATLAS, CMS and LHCb, and the first observation in a single experiment of $D^0-\overline{D}^0$ mixing   from LHCb, an impressive $9.1\,\sigma$. LHCb also presented new world-best $B^0$ and $B_s$ mixing measurements, shown in Fig.\,\ref{Fig:mixing}, giving $\Delta m_d = (0.5156 \pm 0.0051  \pm 0.0033)$\,ps$^{-1}$\,\cite{LHCbbmixing} and $\Delta m_s = (17.768 \pm 0.023  \pm 0.006)$\,ps$^{-1}$\,\cite{LHCbbsmixing}, respectively.

Beauty 2014 in Edinburgh\,\cite{Beauty2014} saw the name of the conference change.  ``International Conference on $B$-Physics at Hadron Machines'' became ``International Conference on $B$-Physics at Frontier Machines'', to better reflect democracy in  results from $pp$ and $p(\overline{p})$  and $e^+e^-$ machines.  Improved $\gamma (\phi_3)$ combination measurements to   $9^\circ$ were reported at the conference, and LHCb first reported an interesting anomaly in the so-called $P_5^\prime$ variable in $B\rightarrow K^*\mu\mu$ decays. In addition, preparation had started on the new Belle-II experiment in Japan.

\begin{figure}[h]
  \begin{center}
    \includegraphics[width=0.75\linewidth]{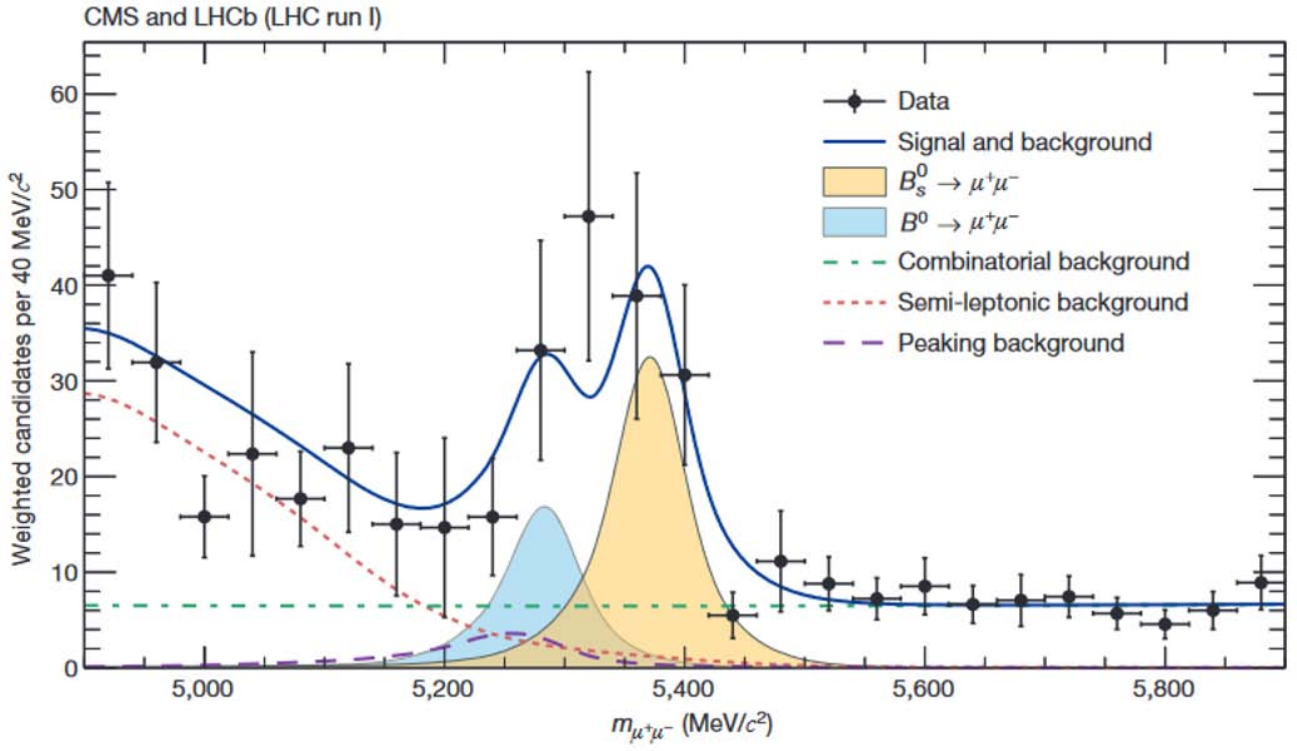} 
  \end{center}
  \caption{
The $\mu^+\mu^-$ mass spectrum from a combination of LHCb and CMS data~\cite{Bsmumu}. 
    }
  \label{Fig:btomumu}
\end{figure}

\begin{figure}[h]
  \begin{center}
    \includegraphics[width=0.6\linewidth]{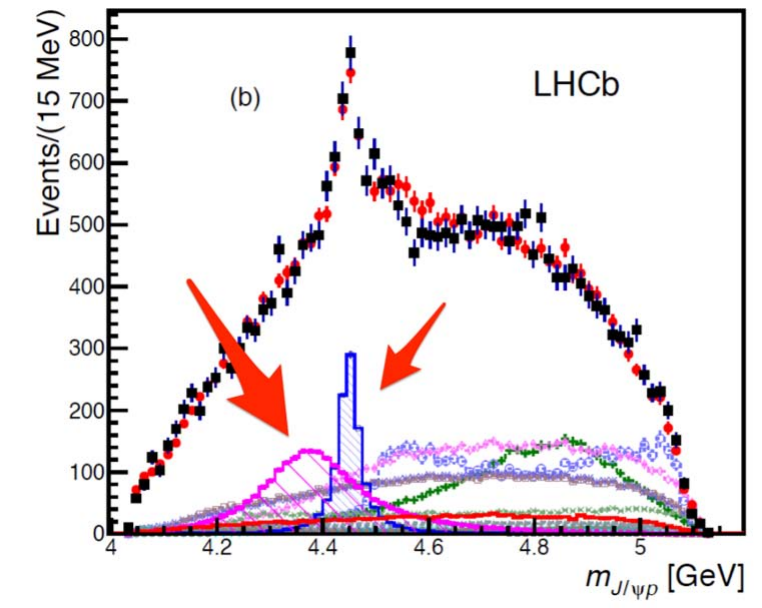} 
  \end{center}
  \caption{
The LHCb measurement of the $(J/\psi p)$ mass spectrum in $\Lambda^0_b\rightarrow K^- J/\psi p$ decays\,\cite{Pentaquark}. The arrows show the $P_c^+(4380)$ and $P_c^+(4450)$ pentaquark states.  }
  \label{Fig:pentaquark}
\end{figure}

The  highlight of Beauty 2016 held in Marseille\,\cite{Beauty2016}  was arguably the first observation of $B_s\rightarrow\mu^+\mu^-$ and evidence for $B^0\rightarrow\mu^+\mu^-$ from a combination of LHCb and CMS measurements~\cite{Bsmumu}, shown in Fig.\,\ref{Fig:btomumu}. This represented a culmination of 35 years of intense searching. The combined fit led to the branching-fraction measurements $\mathcal{B}(B_s\rightarrow\mu^+\mu^-)= (2.8\substack{+0.7 \\ -0.6})\times 10^{-9}$ and $\mathcal{B}(B_s\rightarrow\mu^+\mu^-)= (3.9\substack{+1.6 \\ -1.4})\times 10^{-10}$. 
The second highlight was the discovery of the pentaquark, reported by LHCb\,\cite{Pentaquark}, and shown in Fig.\,\ref{Fig:pentaquark}. Two states, the so-called $P_c^+(4380)$ and $P_c^+(4450)$, were observed in the $(J/\psi p)$ mass spectrum in $\Lambda^0_b\rightarrow K^- J/\psi p$ decays, with significances of 9 and 12\,$\sigma$, respectively. A subsequent analysis with more data has shown this system to be even richer than first thought\,\cite{Pentaquark2}.

Beauty 2018 in La Biodola, Isle of Elba\,\cite{Beauty2018}, was the ``year of the anomaly''. This was the first presentation of the LHCb measurements of the quantities $R_K$ and $R_{K^*}$, the ratios of $B$ mesons decaying into $K^{(*)}\mu^+\mu^-$ and $K^{(*)}e^+e^-$. The measurements  differed from unity by $2.5\,\sigma$, potentially hinting at lepton non-universality. This resulted in much  discussion and interesting speculation at the conference. In other news, LHCb measured $\gamma (\phi_3)$ via a combination of decay channels to almost $5^\circ$, namely $\gamma = (74.0\substack{+5.0 \\ -5.8})^\circ$.

A highlight of the Beauty 2019 in Ljubljana, Slovenia\,\cite{Beauty2019}, was LHCb's discovery  of CP violation in the charm system at $5.3\,\sigma$ significance\,\cite{CPcharm}.
The measure of the difference of the CP asymetries in the $D^0$ meson decaying into $K^+K^-$ and $\pi^+\pi^-$, $\Delta A_{CP}$  was measured to be  $(-15.4 \pm 2.9) \times 10^{-4}$.
The Belle-II collaboration was congratulated on the  first data-taking run of the experiment.

\section{The era of Belle-II and the LHC upgrades}


The year 2020 saw the Covid pandemic and Beauty 2020, hosted by Tokyo,  was the first completely online Beauty conference\,\cite{Beauty2020}. Although participants were unable to meet in person,   many exciting results were presented. The conference highlighted first results from Belle-II, an example being the mass distribution and angular distribution of $B^+\rightarrow\phi K^+$.

The current meeting, Beauty 2023 in Clermont-Ferrand, saw the LHC resuming  operation after three years of long shut-down and the LHCb Upgrade-I start  taking data. Together with Belle-II, a new fresh exciting chapter for flavour physics had started. 
The community can look forward to the ATLAS and CMS upgrades in Long Shutdown 3 of the LHC and the LHCb Upgrade-II in the early 2030's, and the high-quality and no-doubt surprising results that will surely come from these projects.

\section{Summary}

\begin{figure}[h]
  \begin{center}
    \includegraphics[width=0.49\linewidth]{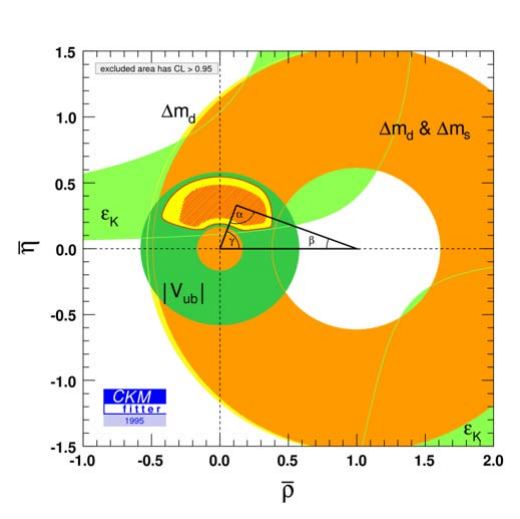} 
        \includegraphics[width=0.49\linewidth]{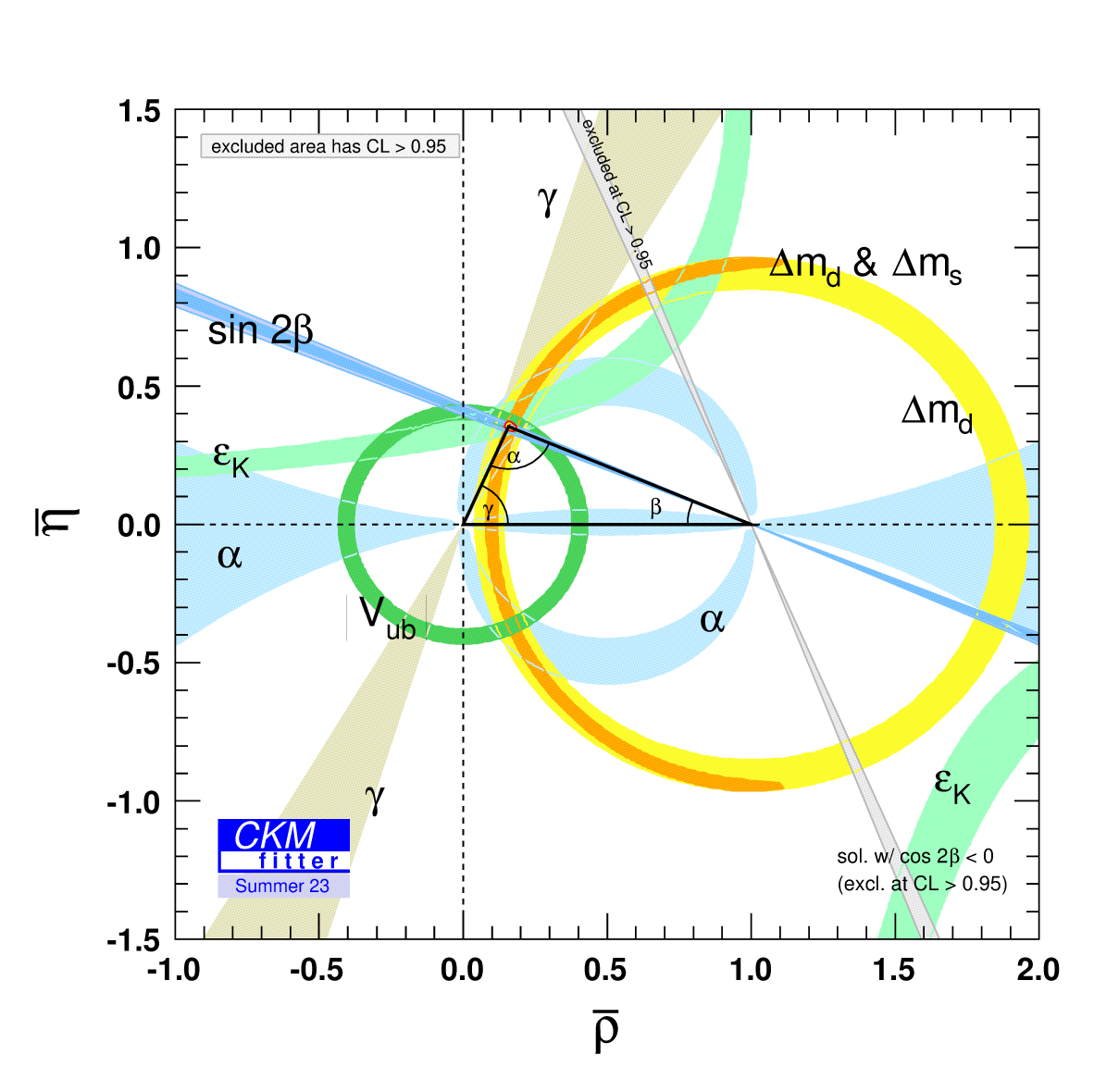} 
  \end{center}
  \caption{
Evolution of the Unitarity Triangle over the 30 years of the Beauty conference\,\cite{CKMfitter}.
    }
  \label{UTevolution}
\end{figure}

The  30-year history of the Beauty conference series   has seen  many exciting experimental  developments, and the  conferences have  been enlightened by a healthy mixture of theoretical presentations.
Figure\,\ref{UTevolution} shows the evolution of the Unitarity Triangle over this period\,~\cite{CKMfitter}.

Over the last thirty years, rare decays have been measured down to the one in a billion level, and whole families of new particle states have been discovered. Unitarity Triangle measurements are consistent with the Standard Model and new physics is becoming increasingly constrained.
Nevertheless there is still need for increased precision, which Belle-II and the LHC upgrades will provide in future years.
Although the field of Flavour Physics has taken a huge leap over the last 30 years, there still remains a whole lot  to learn. 
We all look forward to the next 30 years of the Beauty conference series!

\section*{Acknowledgments}
I would like to thank the organisers of  Beauty 2023  for making the conference so enjoyable. In particular, I would like to thank the conference chairs  Robert Fleischer and Guy Wilkinson, and the hospitable and efficient local team lead by St$\acute{{\rm e}}$phane Monteil and R$\acute{{\rm e}}$gis Lefevre. 
Finally, it has been a pleasure to work as as co-chair with Peter Schlein, Samim Erhan and Robert Fleischer for over a decade of the conference.

\end{document}